
\input phyzzx

\rightline{OU-HET 171}
\rightline{November, 1992}
\vskip3.0cm
\centerline{\bf Unification of Gravity, Gauge and Higgs Fields}
\centerline{\bf by Confined Quantum Fields II}
\centerline{\bf -Effective Theory-}
\vskip2.0cm
\centerline{Toshiki Isse\footnote*{Email address
402g0027@ex.ecip.osaka-u.ac.jp}}

\vskip2.0cm

\centerline {\it Department of Physics}
\centerline {\it Osaka University, Toyonaka, Osaka 560, Japan}

\vskip3cm

  Dynamics of quantized free fields ( of spin 0 and 1/2 ) contained in a
subspace $V_*$ of an N+4 dimensional flat space $V$ is studied. The space
$V_*$ is considered as a neighborhood of a four dimensional submanifold $M$
arbitrarily embedded into $V$. We show that Einstein SO(N)-Yang-Mills Higgs
theory is induced as a low energy effective theory of the system. Gravity,
SO(N) gauge fields and Higgs fields are obtained from embedding functions
of $M$.
\medskip
\vfill
\eject

\beginsection 1 Introduction

  In a previous paper [1], we discussed classical dynamics of free fields
that are contained in a subspace $V_*$ of an N+4 dimensional flat space
$V$. A metric of $V$ is $\eta _{\alpha \beta }\dot =(-1,1,...,1)$ $(\alpha
,\beta =0,1,...,N+3)$. We regard $V_*$ as a neighborhood of a four
dimensional submanifold $M$ arbitrarily embedded into $V$. In other words,
we assume an existence of some physical object (we call it
$B
spacetime) which confines quantized free fields inside the neighborhood of
$M$. ( See Fig[1].)

FIG[1]

In the case of spin 1/2 fields, the system was shown to be described by an
infinite number of fields in $M$, interacting with gravity ($g_{\mu \nu
}$), SO(N) gauge fields ($A_\mu $) and an N-plet of scalar fields ($\phi
^a$). The fields $g_{\mu \nu }$, $A_\mu $and $\phi ^a$ are determined from
embedding functions of $M$ and correspond respectively to the induced
metric, the normal connection and the extrinsic curvature of $M$.
 In this paper, we study low energy effective theory of the system.
Investigating quantum effects caused by the massive modes of the fields we
will show that Einstein-SO(N)-Yang-Mills-Higgs theory is induced as a low
energy effective theory of the system. The fields $\phi ^a$ will be shown
to behave as Higgs fields.

 \beginsection 2 Effective action for scalar field

 We first review classical dynamics of a free massless scalar field that is
contained in the neighborhood $V_*$ of $M$. The Neumann boundary condition
is imposed on a boundary of $V_*$. The action of the system is described by
$$
S_{scalar}[D,X]=\int\limits_{V_*}^{} {dZ^{N+4}(-\eta ^{\alpha \beta
}{{\partial D^*(Z)} \over {\partial Z^\alpha }}{{\partial D(Z)} \over
{\partial Z^\beta }}+constant)},\eqno(2.1)
$$
  (The arbitrary constant introduced in eq.(2.1) becomes a cosmological
constant in eq.(2.5). ) The exact definition of $V_*$ will be given
shortly. We find that the appropriate coordinates to describe $V_*$ are
given by the curvilinear coordinates $(x_\mu ,\varsigma ^a)$, where $x_\mu
(\mu =0,1,2,3)$ are the coordinates tangent to $M$ and $\varsigma ^a(\
a=1,...,N)$ are the ones normal to $M$. The line element of $V$ is written
in the new coordinates,
$$
ds^2=\sum\limits_{\mu ,\nu =0}^3 {g_{\mu \nu }(x,\varsigma )dx^\mu dx^\nu
}+\sum\limits_{a=1}^N {d\varsigma ^ad\varsigma ^a}.\eqno(2.2)
$$
   We can choose $\varsigma ^a=0$ on the manifold $M$, so that $x_\mu (\mu
=0,1,2,3)$ can be regarded as the coordinates of $M$. Using the new
coordinates, we define the neighborhood $V_*$ as a region that satisfies
$\left| {\varsigma ^a} \right|\le l/2,\  (a=1,...,N)$. We can always find
the above coordinates if all focal points lie outside the neighborhood of
$M$. In order to satisfy the above condition, $M$ has to be embedded so
that the smallest focal length $d(X)$ is larger than $\ell /2$.
  In the new coordinates, the Lagrangian of the system is written by
$$
\eqalign{&S_{scalar}[D,X]=\int\limits_{V_*}^{} {\sqrt
{-g(X(x))}dx^4d\varsigma ^N\ [g^{\mu \nu }(X(x))(1+\Delta ){{\partial
D^*(x,\varsigma )} \over {\partial x^\mu }}{{\partial D(x,\varsigma )}
\over {\partial x^\nu }}}\cr
  &\                     \ \ \ +(1+\Delta '){{\partial D^*(x,\varsigma )}
\over {\partial \varsigma ^b}}{{\partial D(x,\varsigma )} \over {\partial
\varsigma ^b}}+constant],\cr}\eqno(2.3a)
$$
where $g_{\mu \nu }(X(x))$ is the induced metric of $M$ and the correction
terms  $\Delta $and $\Delta '$ are order of magnitude [1]
$$
\Delta \cong \Delta '\cong O({\ell  \over {d(X)}}).\eqno(2.3b)
$$
 We neglect the correction terms $\Delta $and $\Delta '$ as we are
interested only in the limit $\ell /d(X)\to 0$. We will discuss the
physical meaning of the limit later. Then we decompose the scalar field
into a Fourier series with respect to $\varsigma ^a$,
$$
D(x^\mu ,\varsigma ^a)=l^{-N/2}\sum\limits_n^{} {D(x^\mu ,n)\exp (i2\pi
n^a\cdot \varsigma ^a/l)}.\eqno(2.4)
$$
 We substitute the above expression into the action (2.3a) and integrate
over the $\varsigma ^a$ variables, neglecting $\Delta $and $\Delta '$. We
find
$$
\eqalign{&S_{scalar}[D,X]\cr
  &=\int\limits_M^{} {dx^4\sqrt {g(x)}[\sum\limits_n^{} {(-g^{\mu \nu
}(x){{\partial D^*(x,n)} \over {\partial x^\mu }}{{\partial D(x,n)} \over
{\partial x^\nu }}+M_n^2\left. {\left| {D(x,n)} \right.} \right|^2}})\cr
  &\                     \ \ \ \ \ \ \ \ \ \ \ \ \ \ \ \
+constant],\cr}\eqno(2.5)
$$
where
$$
M_n^2=({{2\pi n} \over \ell })^2.\eqno(2.6)
$$
 The dynamics of the system obtained as the action of four dimensional
spacetime is nothing but an infinite number of scalar fields in curved four
dimensional space. Different modes would couple with one another in the
action (2.5), if the correction terms $\Delta $and $\Delta '$are present.
  The mass of $n$($\ne 0$) modes are very large, when $l$ is very small.
The  massive modes  cannot be excited classically in low energy physics.
Only the zero modes remain in  eq. (2.5). We find, however, that quantum
effects caused by the massive modes are not negligible. The massive modes
of the fields are influenced by the configuration of the submanifold, which
itself provides the gravity. It is therefore natural to expect that the
quantum effects caused by the massive modes generate dynamics of the
gravity. Such an idea was first introduced by Sakharov and Zel$B
[2][3], who proposed that  Einstein action is not a fundamental microscopic
action but rather an effective action induced by vacuum quantum structure.
( We also explain the situation in other words. The heavy particles bounce
off the  $B
dynamics of the template of spacetime, which can be interpreted as if the
gravity has dynamics. )

 The effective Lagrangian of the system is obtained by path integrating all
massive modes. As the different modes in (2.5) are not coupled to one
another, it sufficient to compute the effective action for a single mode
and to sum overall modes. We write the effective Lagrangian of the system
as
$$
L_{total\ eff}^{(0)}=L_{zero-mode}^{(0)}+L_{eff}^{(0)}+\sqrt
{-g(x)}constant,\eqno(2.7)
$$
where $L_{eff}^{(0)}(g)$ is given as follows,
$$
\eqalign{&L_{eff}^{(0)}=\sum\limits_{n\ne 0}^{} {iTr\ln [-\nabla _\mu
\nabla ^\mu +M_n]}\cr
  &\       =\sum\limits_{n\ne 0}^{} {L_{eff}^{(0)}(n)}.\cr}\eqno(2.8)
$$

 As the gravitational fields are background fields and $D$ is a bilinear
field, it is sufficient to compute one-loop effective action of mode $n$ ;
$L_{eff}^{(0)}(n)$. We evaluate $L_{eff}^{(0)}(n)$ by using a heat kernel
technique developed by De-Witt and Schwinger[4][5],
$$
L_{eff}^{(0)}(n)=i\int_0^\infty  {{{g^{1/2}e^{-iM^2_ns}} \over {(4\pi
is)^{2+\varepsilon /2}}}}\sum\limits_{r=0}^\infty  {a_r(is)^r{{ds} \over
{is}}}.\eqno(2.9)
$$
   Here we used a dimensional regularization n=4+e, where n is taken to be
the dimension of the submanifold $M$. The De-Witt Schwinger coefficients
$a_r$ are calculated as
$$
a_0=1,\eqno(2.10)
$$
$$
a_1={R \over 6},\eqno(2.11)
$$
$$
a_2={{R_{;\mu }^\mu } \over {30}}+{{R^2} \over {72}}-{{R_{\mu \nu }R^{\mu
\nu }} \over {180}}+{{R_{\mu \nu \lambda \rho }R^{\mu \nu \lambda \rho }}
\over {180}},\eqno(2.12)
$$
$$
...
$$
 Carrying out the $s$ integration we obtain
$$
L_{eff}^{(0)}(n)={{g^{1/2}} \over {(4\pi )^2}}\sum\limits_{r=0}^\infty
{\Gamma (r-2-\varepsilon /2){({{2\pi } \over \ell
})^2}n^2}^{-r+2+\varepsilon /2}a_r.\eqno(2.13)
$$
 The mode sum of the effective action can be calculated by using Epstein
zeta functions;
$$\varsigma _N(a_1,...,a_N;s)=\sum\limits_{n_1=-\infty }^\infty  {\cdot
\cdot \cdot }\sum\limits_{n_N=-\infty }^{\infty \  ,} {[(a_1n_1)^2+\cdot
\cdot \cdot (a_Nn_N)^2]^{-s/2}},\eqno(2.14)$$
$$
\varsigma _N(s)\equiv \varsigma _N(1,...,1;s)\eqno(2.15),
$$
where the prime indicates that we should omit the term for which all
${n_i=0}$. These functions satisfy the reflection formula [6][7],
$$
\eqalign{&\Gamma (s/2)\pi ^{-s/2}\varsigma _N(a_1,...,a_N;s)\cr
  &=a_1^{-1}\cdot \cdot \cdot a_N^{-1}\Gamma ([p-s]/2)\pi
^{(s-p)/2}\varsigma _N(1/a_1,...,1/a_N;p-s).\cr}\eqno(2.16)
$$
 The final expression for $L_{eff}^{(0)}$ is
$$
\eqalign{&L_{eff}^{(0)}=\sqrt {g(X(x))}\left( {{{\Gamma (N/2+2)\varsigma
_N(N+4)} \over {\pi ^{N/2+2}\ell ^4}}+{{\Gamma (N/2+1)\varsigma _N(N+2)}
\over {24\pi ^{N/2+2}\ell ^2}}R} \right.\cr
  &\  \left. {\     +{1 \over {8\pi ^2\varepsilon }}[{{R^2} \over
{72}}+{{R^{\mu \nu \lambda \rho }R_{\mu \nu \lambda \rho }} \over
{180}}-{{R^{\mu \nu }R_{\mu \nu }} \over {180}}]+({\ell  \over {2\pi
}})^2{1 \over \varepsilon }[higher\ order\ terms]} \right)\cr}\eqno(2.17)
$$
where higher order terms are given by the following finite terms,
$$
{1 \over {(4\pi )^2}}\sum\limits_{r=1}^\infty  {\Gamma (r)\varepsilon
\varsigma _N(2r-\varepsilon )({\ell  \over {2\pi
}})^{2r-2}}a_{r+2}.\eqno(2.18)
$$
 It is known that $\varsigma _N(s)$ is convergent when $s>N$ and that it
has simple poles at $s=N,N-1,...,1$[7]. Therefore the coefficients $a_0$
and $a_1$ are finite while $a_r$ ( $[(N+5)/2]\ge $r $\ge $2 ) are divergent
as $\varepsilon \to 0$.

\beginsection 3 Effective action for spinor fields

  In this section we discuss the spinor case. The action of free spinor
fields contained in the neighborhood $V_*$ of $M$ is,
$$
S_{spinor}[\Phi ,X]={1 \over 2}\int\limits_{\  V_*}^{} {dZ^{N+4}[i\bar \Phi
(Z)\Gamma ^\alpha ({{\vec \partial } \over {\partial Z^\alpha
}}-{{\mathord{\buildrel{\lower3pt\hbox{$\scriptscriptstyle\rightarrow$}}\ove
r \partial } } \over {\partial Z^\alpha }})\Phi (Z)+constant]}.\eqno(3.1)
$$
  As in the case of the scalar fields we write the action in the
coordinates ${x_\mu ,\varsigma ^a}$. Then we decompose the spinor fields
into a Fourier series with respect to $\varsigma ^a$ and integrate over
$\varsigma ^a$ in the limit $\ell /d(X)\to 0$. We find the action in the
limit;
$$
\eqalign{&S_{spinor}[\Phi ,X]\cr
  &=\int\limits_{\  M}^{} {\sqrt {-g(x)}dx^4\left( {{i \over 2}\bar \Phi
(x,n)\Gamma ^\mu (x){{\partial \Phi (x,n)} \over {\partial x^\mu }}}
\right.-{i \over 2}{{\partial \bar \Phi (x,n)} \over {\partial x^\mu
}}\Gamma ^\mu (x)\Phi (x,n)}\cr
  &\  +{{i2\pi n^a} \over \ell }\bar \Phi (x,n)\left. {\Gamma ^\alpha
{{\partial \varsigma ^a} \over {\partial Z^\alpha }}} \right|_{\varsigma
=0}\Phi (x,n)+constant\left. {^{^{^{^{}}}}} \right),\cr}\eqno(3.2)
$$
where $\Phi (x^\mu ,n)$ are Fourier components of the spinor fields and
$$
\Gamma ^\mu (x)\equiv \sum\limits_{\alpha =0}^{N+3} {\left. {{{\partial
x^\mu } \over {\partial Z^\alpha }}} \right|_{\varsigma =0}\Gamma ^\alpha
}.\eqno(3.3)
$$
 We have to perform a local transformation on the fields $\Phi (x,n)$to
make fields transforming as spinors on $M$. The transformation is given as
follows
$$
\Phi (x,n)\to \Psi (x,n)=\rho ^{-1}(e(x))\Phi (x,n),\eqno(3.4)
$$
where $e(x)$ is an SO(1,N+3) matrix field on $M$ and $\rho $ is a spinor
representation of $e(x)$. The field  $e(x)$ is defined by putting N+4
column vectors $e_i(x)$ (i=0,..., N+3) on $M$ successively
$$
e(x)=\left[ {e_0(x)...e_3(x)e_4(x)...e_{3+N}(x)} \right].\eqno(3.5)
$$
 \ Here $e_0(x),...,e_3(x)$ are orthonormal vectors (in $V$) tangent to $M$
and $e_4(x),$...,
$e_{3+N}(x)$ are othonormal vectors normal to $M$ [8]. The action (3.2) is
written in terms of $\Psi (x,n)$
$$
\eqalign{&S_{spinor}[\Phi ,X]\cr
  &=\int\limits_M^{} {\sqrt {-g(x)}dx^4\ \left( {i\sum\limits_n^{} {\bar
\Psi (x,n)[\gamma ^\mu (x)\vec \nabla _\mu }+im_n]\Psi (x,n)+constant}
\right),}\cr}\eqno(3.6)
$$
where
$$
\vec \nabla _\mu \equiv {{\vec \partial } \over {\partial x^\mu }}+\omega
_\mu ^{lm}(e(x)){{[\gamma _l,\gamma _m]} \over 8}+A_\mu ^{ab}(e(x)){{[\hat
\Gamma _a,\hat \Gamma _b]} \over 8},\eqno(3.7a)$$
$$
\omega _{\mu \  m}^{\  l}(e(x))=\sum\limits_{\alpha =0}^{N+3}
{[e^{-1}(x)]_\alpha ^l}{{\partial [e(x)]^\alpha _m} \over {\partial x^\mu
}},\eqno(3.7b)$$
$$
A_{\mu \  b}^{\  a}(e(x))=\sum\limits_{\alpha =0}^{N+3} {[e^{-1}(x)]_\alpha
^{a+3}}{{\partial [e(x)]^\alpha _{b+3}} \over {\partial x^\mu
}},\eqno(3.7c)$$
$$
m_n={{2\pi n^a\Gamma ^{3+a}} \over \ell }.\eqno(3.7c)$$
   We showed in Ref.[1] that $\omega _\mu $ and $A_\mu $ are identified as
a spin connection and an SO(N) gauge field respectively and that $\Psi
(x,n)$ transforms as a $2^{[N/2]}$-plet of Dirac spinors on $M$. A local
Lorentz ( a gauge ) transformation corresponds to a transformation of the
orthonormal vectors $e_0(x),...,e_3(x)$ ( $e_4(x),...,e_{3+N}(x)$). The
fields $\omega _\mu $ and $A_\mu $ are determined by the embedding
functions of $M$ modulo the Local Lorentz and the gauge transformations.
Note that $\omega _\mu $ and $A_\mu $ are not independent and that they are
both nontrivial connections.
  In addition to the action (3.1) we consider the following surface term ,
$$
S_{surface }=\int_{V_*} {dZ^{N+4}}h\bar \Phi \Gamma ^\alpha ({{\vec
\partial } \over {\partial Z^\alpha
}}+{{\mathord{\buildrel{\lower3pt\hbox{$\scriptscriptstyle\rightarrow$}}\ove
r \partial } } \over {\partial Z^\alpha }}),\eqno(3.8)
$$
where $h$ is an arbitrary constant. Following the same procedure as
discussed above, we find that the surface term (3.8) induces a simple
Yukawa interaction and an N-plet of scalar fields,
$$
S_{surface }=\int\limits_M {dx^4\sqrt {-g(x)}}h\bar \Psi (x,n)\not \phi
\Psi (x,n).\eqno(3.9)
$$
where
$$\not \phi =\phi ^a\Gamma ^{3+a},\eqno(3.10)
$$
$$
\phi ^a\equiv \sum\limits_{l=0}^3 {\phi _l^{l\  3+a}}.\eqno(3.11)
$$
 The field $\phi _\mu ^{l\  3+a}$ is called extrinsic curvature of $M$ and
is given by $e(x)$ as
$$
\phi _\mu ^{l\  3+a}=\sum\limits_{\alpha =0}^{N+3} {[e^{-1}(x)]_\alpha
^l{{\partial [e(x)]^{\alpha 3+a}} \over {\partial x^\mu }}}.\eqno(3.12)
$$
We find that the final expression for the action of the system in the limit
$\ell /d(X)\to 0$ is
$$
\eqalign{&S_{spinor}[\Phi ,X]\cr
  &=\int\limits_M^{} {\sqrt {-g(x)}dx^4[\ \sum\limits_n^{} {\bar \Psi
(x,n)(i\gamma ^\mu (x)\vec \nabla _\mu }+h\not \phi -m_n)\Psi
(x,n)+}constant].\cr}\eqno(3.13)
$$
  As in the case of the scalar, an effective Lagrangian of the system is
obtained by path integrating all massive modes,
$$
L_{total\ eff}^{(1/2)}=L_{zero-mode}^{(1/2)}+L_{eff}^{(1/2)}+\sqrt
{-g(x)}constant,\eqno(3.14)
$$
where
$$
\eqalign{&L_{eff}^{(1/2)}=-\sum\limits_{n\ne 0}^{} {iTr\ln [i\gamma ^\mu
\nabla _\mu +(h\not \phi -m_n)]}\cr
  &\        =\sum\limits_{n\ne 0}^{} {L_{eff}^{(1/2)}(n)}.\cr}\eqno(3.15)
$$
 The effective action of mode $n$ ; $L_{eff}^{(1/2)}(n)$ is written as
$$
L_{eff}^{(1/2)}(n)=-{i \over 2}Tr\ln (-\nabla _\mu \nabla ^\mu
+Q_n+M^2_n),\eqno(3.16)
$$
where the matrix valued functions $Q_n$ are defined as follows,
$$
Q_n={R \over 4}-{1 \over 2}\gamma ^\mu \gamma ^\nu F_{\mu \nu }+ih\gamma
^\mu (\nabla _\mu \not \phi )+h^2\phi ^a\phi _a-{{4\pi h} \over \ell
}n_a\phi ^a.\eqno(3.17)
$$
  The De-Witt Schwinger expansion of eq.(3.16) is given by
$$
L_{eff}^{(1/2)}(n)=-{i \over 2}\int_0^\infty  {{{g^{1/2}e^{-iM^2_ns}} \over
{(4\pi is)^{2+\varepsilon /2}}}}\sum\limits_{r=0}^\infty
{Tra_r^{(1/2)}(is)^r{{ds} \over {is}}}.\eqno(3.18)
$$
   Here the coefficients $a_r^{(1/2)}$ are
$$
a_0^{(1/2)}=1,\eqno(3.19)
$$
$$
a_1^{(1/2)}={R \over 6}-Q_n,\eqno(3.20)
$$
$$
a_2^{(1/2)}={1 \over {12}}W_{\mu \nu }W^{\mu \nu }+{1 \over 2}Q_n^2-{{RQ_n}
\over 6},...\eqno(3.21)
$$
where
$$
W_{\mu \nu }={1 \over 8}R_{\mu \nu \lambda \rho }[\gamma ^\lambda ,\gamma
^\rho ]+F_{\mu \nu }.\eqno(3.22)
$$
 A mode sum is performed by using the Epstein zeta functions,
$$
\eqalign{&L_{eff}^{(1/2)}=\sqrt {-g}\left( {\ell ^{-4}c_1} \right.+\ell
^{-2}[c_2R+c_3\phi ^2]\cr
  &\             -{{N_F} \over {(4\pi )^2\varepsilon }}[{{R^2} \over
{72}}-{{R_{\mu \nu }R^{\mu \nu }} \over {45}}-{{7R_{\mu \nu \lambda \rho
}R^{\mu \nu \lambda \rho }} \over {360}}\cr}$$
$$\eqalign{&\            -{2 \over {3N_F}}trF_{\mu \nu }F^{\mu \nu
}+2h^2(\nabla _\mu \phi )^a(\nabla _\mu \phi )_a+2h^4(\phi ^2)^2\cr
  &\             +{1 \over 3}h^2\phi ^2R]+{{\ell ^2} \over \varepsilon
}[higer\ order\ terms]\left. {} \right).\cr}\eqno(3.23)
$$
 Here
$$
c_1=-{{2N_F\Gamma (N/2+2)\varsigma _N(N+4)} \over {\pi
^{N/2+2}}}\eqno(3.24a)$$

$$c_2={{N_F\Gamma (N/2+1)\varsigma _N(N+2)} \over {24\pi
^{N/2+2}}}\eqno(3.24b)$$

$$c_3={{(1+2/N)N_F\Gamma (N/2+1)\varsigma _N(N+2)h^2} \over {2\pi
^{N/2+2}}},\eqno(3.24c)
$$
and  $B
$a_r^{(1/2)}(r\ge 3)$.

\beginsection 4  Interpretation of effective action

  In quantum field theory divergent terms are renormalized by redefining
coupling constants and fields. We cannot, however, remove the divergent
terms in eq.(2.17) and eq.(3.23) by usual renormalization prescription;
coupling constants to be renormalized do not exist in our original
Lagrangians (2.1),(3.1). ( The arbitrary constants appearing in eq.(2.1)
and eq.(3.1) are  parameters which renormalize the cosmological constant. )
Is it a difficulty in our theory? The answer is no. The theory we developed
here is nothing but a free field theory restricted in a finite region $V_*$
of the embedding space $V$. Therefore we must obtain some reasonable
physical results from eq.(2.17) and eq.(3.23).
 Although we cannot renormalize the effective actions (2.17) and (3.23), we
will propose a possible effective Lagrangian which describe quantum
fluctuation of the system. We would like to search for a physically
reasonable effective Lagrangian by the following method. We assume that
there exist finite physical fields denoted by  $\tilde X^\alpha (x)$,
$\tilde \Psi (x,0)$ and $\tilde D(x,0)$, which are defined respectively by
rescaling the original fields
$$
X^\alpha (x)=\varepsilon ^{-\lambda /2}\tilde X^\alpha (x)\eqno(4.1a)
$$
$$
\Psi (x,0)=(N_F/6\pi ^2)^{1/2}\kappa _\varphi \varepsilon ^{-\lambda
_\varphi }\tilde \Psi (x,0)\eqno(4.1b)
$$
$$
D(x,0)=(N_F/6\pi ^2)^{1/2}\kappa _D\varepsilon ^{-\lambda _D}\tilde
D(x,0).\eqno(4.1c)
$$
where $\kappa _\varphi $ and $\kappa _D$ are arbitrary constants. ( The
coefficients in front of $\kappa _\varphi $ and $\kappa _D$ are purely
conventions. ) The parameters $\lambda $, $\lambda _\varphi $, and $\lambda
_D$ are introduced as powers of $\varepsilon $ multiplying the original
fields. In other words, we regard $X^\alpha (x)$, $\Psi (x,0)$ and $D(x,0)$
as divergent fields. Eq. (4.1a) implies
$$
g_{\mu \nu }(X(x))=\varepsilon ^{-\lambda }\tilde g_{\mu \nu }(\tilde
X(x))\eqno(4.2)$$
$$
e_\mu ^l=\varepsilon ^{-\lambda /2}\tilde e_\mu ^l.\eqno(4.3)
$$
  Consider a corresponding system that consists of both spinor fields and
$N_B$ scalar fields. An effective action of the system is written in terms
of the new fields;
$$\eqalign{&L_{eff}^{}=\sqrt {-\tilde g}(\sum\limits_{A=1}^{N_B} {\partial
_\mu \tilde D^{*(A)}(x,0)\partial ^\mu \tilde D^{*(A)}(x,0)}\kappa
_D^2\varepsilon ^{-\lambda -2\lambda _D}{{N_F} \over {6\pi ^2}}\cr
  &\           -\bar {\tilde \Psi} (x,0)(i\tilde \gamma ^\mu \nabla _\mu
+h\tilde \phi ^a\Gamma ^{a+3})\tilde \Psi (x,0)\kappa _\varphi
^2\varepsilon ^{-3\lambda /2-2\lambda _\varphi }{{N_F} \over {6\pi ^2}}\cr
  &\           +constant+d_1\tilde R\varepsilon ^{-\lambda }\cr
  &\           +{{N_F} \over {8\pi ^2\varepsilon }}[({{N_B} \over {N_F}}-{1
\over 2}){{\tilde R^2} \over {72}}+(-{{N_B} \over {N_F}}+2){{\tilde R_{\mu
\nu }\tilde R^{\mu \nu }} \over {180}}+({{4N_B} \over {N_F}}+7){{\tilde
R_{\mu \nu \lambda \rho }\tilde R^{\mu \nu \lambda \rho }} \over {720}}\cr
  &\           -\sum\limits_{c=1}^{\dim SO(N)} {{{N_FF^{(c)}_{\mu \nu
}F^{(c)\mu \nu }} \over {24\pi ^2\varepsilon }}}-{{N_Fh^2} \over {8\pi
^2\varepsilon }}[{1 \over 2}(\nabla _\mu \tilde \phi )^2+h^2(\tilde \phi
^2)^2]+d_2\tilde \phi ^2\varepsilon ^{-\lambda }\cr
  &\           -{{N_Fh^2\tilde \phi ^2\tilde R} \over {48\pi ^2\varepsilon
}}+\ell ^2\varepsilon ^{3\lambda -1}[higher\ order\
terms]),\cr}\eqno(4.4)$$
where spacetime indices are contracted by using $\tilde g^{\mu \nu }$ and
generators  $T^a$
 of the SO(N) group in the spinor representation are normalized as
$trT^aT^b$$=$$-\delta ^{ab}N_F/4$  (  a=1,...,dimSO(N)). The fields $\tilde
\phi ^a$, the matrices $\tilde \gamma ^\mu $ and the constants $d_1$ and
$d_2$ in the above equation are respectively defined as
$$
\tilde \phi ^a\equiv \sum\limits_{l.\mu =0}^3 {\tilde e^\mu _l\phi _\mu
^{l\  3+a}}\eqno(4.5)$$
$$
\tilde \gamma ^\mu \equiv \sum\limits_{l=0}^3 {\tilde e^\mu _l\gamma
^l}\eqno(4.6)$$
$$
d_1={{\Gamma (N/2+1)\varsigma _N(N+2)(N_B+N_F)} \over {24\pi ^{N/2+2}\ell
^2}}\eqno(4.7)$$
$$
d_2={{(1/2+1/N)\Gamma (N/2+1)\varsigma _N(N+2)N_F} \over {24\pi
^{N/2+2}\ell ^2}}.\eqno(4.8)$$
Taking into account the fact that the fields $\tilde X^\alpha (x)$, $\tilde
\Psi (x,0)$ and $\tilde D(x,0)$ are finite, we factorize a leading
divergent coefficient as an overall factor of the Lagrangian (4.4). For
example, when $\lambda =2$ ( $\lambda _D=(1-\lambda )/2$,$\lambda _\varphi
=(1-3\lambda /2)/2$ ), the effective action (4.4) is written as
$$
L_{eff}={{\kappa ^2_\varphi N_F} \over {6\pi ^2\varepsilon
}}L_{eff}^{phys}\eqno(4.9)
$$
 Here  $L_{eff}^{phys}$ is a finite Lagrangian of the physical fields,
$$\eqalign{&L_{eff}^{phys}\cr
  &=\sqrt {-\tilde g}[-\sum\limits_{A=1}^{N_B} {\partial _\mu \tilde
D^{*(A)}(x,0)\partial ^\mu \tilde D^{*(A)}(x,0)}\cr
  &-\bar{\tilde \Psi} (x,0)(i\tilde \gamma ^\mu \nabla _\mu +\kappa
_\varphi \hat \phi ^a\Gamma ^{a+3})\tilde \Psi (x,0)+contant+{1 \over
{16\pi G}}\tilde R\cr
  &\   -\sum\limits_{c=1}^{\dim SO(N)} {{{F^{(c)}_{\mu \nu }F^{(c)\mu \nu
}} \over {4\kappa _\varphi ^2}}}-{1 \over 2}(\nabla _\mu \hat \phi
)^2+{{m^2} \over 2}\hat \phi ^2-{{\kappa _\varphi ^2} \over 3}(\hat \phi
^2)^2-{{\hat \phi ^2\tilde R} \over {8\kappa _\varphi ^2}}\cr
  &+{3 \over {4\kappa _\varphi ^2}}[({{N_B} \over {N_F}}-{1 \over
2}){{\tilde R^2} \over {72}}+(-{{N_B} \over {N_F}}+2){{\tilde R_{\mu \nu
}\tilde R^{\mu \nu }} \over {180}}+({{4N_B} \over {N_F}}+7){{\tilde R_{\mu
\nu \lambda \rho }\tilde R^{\mu \nu \lambda \rho }} \over
{720}}]\cr}\eqno(4.10)
$$
where
$$
{1 \over {16\pi G}}={{\Gamma (N/2+2)\varsigma _N(N+4)} \over {4\kappa
_\varphi ^2\pi ^{N/2}\ell ^2}}({{N_B} \over {N_F}}+1)\eqno(4.11)$$
$$
m^2={{(1+2/N)\Gamma (N/2+1)\varsigma _N(N+2)} \over {\kappa _\varphi ^2\pi
^{N/2}\ell ^2}}\eqno(4.12)$$
$$
\hat \phi ^a\equiv {h \over {\kappa _\varphi ^{}}}\sqrt {3/2}\tilde \phi
^a.\eqno(4.13)$$
  As the overall factor of the Lagrangian (4.9) does not effect equation of
motion, we can regard $L_{eff}^{phys}$ as a possible effective Lagrangian
which describes the effective theory of the system. Of course the classical
Lagrangian obtained here depends on the choice of the definition of the
physical fields $\tilde X^\alpha (x)$, $\tilde \Psi (x,0)$ and $\tilde
D(x,0)$, namely the choice of the values $\lambda $, $\lambda _\varphi $,
and $\lambda _D$. All possible effective Lagrangians are classified by
$\lambda $, $\lambda _\varphi $, and $\lambda _D$. We find a classification
with respect to $\lambda $ is physicaly the most relevant. A table [1]
provides the classification with respect to $\lambda$.

Table 1

Remember the condition of the embedding; $d(X)>\ell /2$ and the correction
terms $\Delta $and $\Delta '$ discussed in section two. They are written in
terms of the finite fields $\tilde X^\alpha $ as
$$
d(\tilde X)>\varepsilon ^{\lambda /2}\ell /2 \eqno(4.14)$$
$$
\Delta \cong \Delta '\cong O(\varepsilon ^{\lambda /2}\ell /d(\tilde X)).
\eqno(4.15)
$$

 If $\lambda >0$, the condition of the embedding is automatically satisfied
and the correction terms approach zero in the limit $\varepsilon \to 0$.
 The table [1] shows that energy scale of the system increases as the value
$\lambda $ decreases. A plausible effective Lagrangian which describes the
system in the lowest energy scale is in a case $\lambda >1$. However in
this case the system is unstable. When $\lambda =1$( in the second lowest
energy scale ) the effective Lagrangian $L_{eff}^{phys}$ includes
Einstein-Yang-Mills-Higgs action. In this case gauge symmetry is broken
spontaneously and the fermions and the gauge fields obtain mass. If
$0<\lambda <1$, the symmetry breakdown does not happen nor the effective
action $L_{eff}^{phys}$  includes Einstein action. If $\lambda <0$, the
calculation of the effective action of the system becomes meaningless. In
this case $\ell /d(X)$, which we neglected, becomes  infinitely large. The
system is described not as a four dimensional effective theory in $M$ but
as an N+4 dimensional free field theory in  $V_*$.

\beginsection 5 Conclusion

  We classified possible effective Lagrangians which describe low energy
effective theory of the system. The effective Lagrangian which describes
the lowest energy scale of the system includes Einstein-Yang-Mills-Higgs
action. Gravity, SO(N) gauge fields and Higgs fields are induced themselves
by embedding functions of $M$. However these three kind of fields are not
independent. It is a future task to build a model that can reproduce the
standard GUT theories of SU(5),SO(10), or $E_6$.

\beginsection ACKNOWLEDGMENTS

 I am very grateful to K. Kikkawa, H. Itoyama, H. Kunitomo and H. Suzuki
for useful discussions and comments.

\beginsection REFERENCES

\item{[1]}T. Isse,   ( hep-th/9211002, OU-HET 170 )
\item{[2]} A. D. Saharov, Sov. Phys. Doklady 12 (1968), 1040
     Ya. B Zel$B
\item{[3]}for review see, S.L. Adller, Rev. Mod. Phys., 54 (1982), 729
\item{[4]}B.S. DeWitt, in Raltivity, groups and topology II, ed. B.S. De
Witt and R. Stora
    ( North-Holland, Amsterdam, 1984 ) p. 393
\item{[5]}P.B. Gilkey, J. Diff. Geom. 10 (1975) 601
\item{[6]}J. Ambjorn and S. Wolfram, Ann. Phys. 147 (1983), 1
\item{[7]}K. Kirsten, J. Math. Phys. 32 (1990) 3008,
\item{[8]} S.Kobayashi. and K.Nomizu, Foundation of Differential Geometry
vol I, II                      ( Interscience Publishers , New York, 1969)

\end